\documentclass[
superscriptaddress,aps,preprintnumbers,amsmath,amssymb,prd,nofootinbib,preprint]{revtex4-1}

\usepackage[dvipdfm]{graphicx}
\usepackage{epstopdf}
\usepackage{dcolumn}
\usepackage{bm}
\usepackage{hyperref}
\usepackage{color}

\begin{document}


\def\a{\alpha}
\def\b{\beta}
\def\c{\varepsilon}
\def\d{\delta}
\def\e{\epsilon}
\def\f{\phi}
\def\g{\gamma}
\def\h{\theta}
\def\k{\kappa}
\def\l{\lambda}
\def\m{\mu}
\def\n{\nu}
\def\p{\psi}
\def\q{\partial}
\def\r{\rho}
\def\s{\sigma}
\def\t{\tau}
\def\u{\upsilon}
\def\v{\varphi}
\def\w{\omega}
\def\x{\xi}
\def\y{\eta}
\def\z{\zeta}
\def\D{\Delta}
\def\G{\Gamma}
\def\H{\Theta}
\def\L{\Lambda}
\def\F{\Phi}
\def\P{\Psi}
\def\S{\Sigma}

\def\o{\over}
\def\beq{\begin{eqnarray}}
\def\eeq{\end{eqnarray}}
\newcommand{\gsim}{ \mathop{}_{\textstyle \sim}^{\textstyle >} }
\newcommand{\lsim}{ \mathop{}_{\textstyle \sim}^{\textstyle <} }
\newcommand{\vev}[1]{ \left\langle {#1} \right\rangle }
\newcommand{\bra}[1]{ \langle {#1} | }
\newcommand{\ket}[1]{ | {#1} \rangle }
\newcommand{\EV}{ {\rm eV} }
\newcommand{\KEV}{ {\rm keV} }
\newcommand{\MEV}{ {\rm MeV} }
\newcommand{\GEV}{ {\rm GeV} }
\newcommand{\TEV}{ {\rm TeV} }
\newcommand{\1}{\mbox{1}\hspace{-0.25em}\mbox{l}}
\def\diag{\mathop{\rm diag}\nolimits}
\def\Spin{\mathop{\rm Spin}}
\def\SO{\mathop{\rm SO}}
\def\O{\mathop{\rm O}}
\def\SU{\mathop{\rm SU}}
\def\U{\mathop{\rm U}}
\def\Sp{\mathop{\rm Sp}}
\def\SL{\mathop{\rm SL}}
\def\tr{\mathop{\rm tr}}

\def\IJMP{Int.~J.~Mod.~Phys. }
\def\MPL{Mod.~Phys.~Lett. }
\def\NP{Nucl.~Phys. }
\def\PL{Phys.~Lett. }
\def\PR{Phys.~Rev. }
\def\PRL{Phys.~Rev.~Lett. }
\def\PTP{Prog.~Theor.~Phys. }
\def\ZP{Z.~Phys. }

\def\dd{\mathrm{d}}
\def\ff{\mathrm{f}}
\def\BH{{\rm BH}}
\def\inf{{\rm inf}}
\def\ev{{\rm evap}}
\def\eq{{\rm eq}}
\def\SM{{\rm sm}}
\def\Mpl{M_{\rm Pl}}
\def\GeV{{\rm GeV}}
\newcommand{\Red}[1]{\textcolor{red}{#1}}

\def\mDM{m_{\rm DM}}
\def\mphi{m_{\phi}}
\def\TeV{{\rm TeV}}
\def\Gphi{\Gamma_\phi}
\def\TR{T_{\rm RH}}
\def\Br{{\rm Br}}
\def\DM{{\rm DM}}
\def\Eth{E_{\rm th}}
\newcommand{\lmk}{\left(}  
\newcommand{\rmk}{\right)}
\newcommand{\lkk}{\left[}  
\newcommand{\rkk}{\right]}
\newcommand{\lhk}{\left \{ }  
\newcommand{\rhk}{\right \} }
\newcommand{\del}{\partial}  
\newcommand{\la}{\left\langle} 
\newcommand{\ra}{\right\rangle}


\title{
Lower bound of the tensor-to-scalar ratio $r\gsim 0.1$ in a nearly quadratic chaotic inflation model in supergravity
}

\author{Keisuke Harigaya}
\affiliation{Kavli IPMU (WPI), TODIAS, The University of Tokyo, Kashiwa, 277-8583, Japan}
\author{Masahiro Kawasaki}
\affiliation{ICRR, The University of Tokyo, Kashiwa, 277-8582, Japan}
\affiliation{Kavli IPMU (WPI), TODIAS, The University of Tokyo, Kashiwa, 277-8583, Japan}
\author{Tsutomu T.~Yanagida}
\affiliation{Kavli IPMU (WPI), TODIAS, The University of Tokyo, Kashiwa, 277-8583, Japan}
\begin{abstract}
We consider an initial condition problem in a nearly quadratic chaotic inflation model in supergravity.
We introduce shift symmetry breaking not only in the superpotential but also in the Kahler potential.
In this model the inflaton potential is nearly quadratic for inflaton field values around the Planck scale, but deviates from the quadratic one for larger field values.
As a result, the prediction on the tensor-to-scalar ratio can be smaller than that of a purely quadratic model. 
Due to the shift symmetry breaking in the Kahler potential, the inflaton potential becomes steep for large inflaton field values,
which may prevent inflation from naturally taking place in a closed universe.
We estimate an upper bound on the magnitude of the shift symmetry breaking so that inflation takes place before a closed universe with a Planck length size collapses, which yields a lower bound on the tensor-to-scalar ratio, $r\gsim 0.1$.
\end{abstract}

\date{\today}
\maketitle
\preprint{IPMU 14-0326}

\section{Introduction}

Cosmic inflation is an essential ingredient of the modern cosmology.
It solves the horizon and flatness problem~\cite{Guth:1980zm,Kazanas:1980tx}.
Further, so-called slow-roll inflation~\cite{Linde:1981mu,Albrecht:1982wi} (see also Ref.~\cite{Starobinsky:1980te})
predicts almost scale invariant cosmic density perturbations~\cite{Mukhanov:1981xt,Hawking:1982cz,Starobinsky:1982ee,Guth:1982ec,Bardeen:1983qw},
which explains the observed large scale structure of the universe and the cosmic microwave background~\cite{Hinshaw:2012aka,Ade:2013zuv}.

Among inflation models, chaotic inflation~\cite{Linde:1983gd} is the most attractive model since it is free from the initial condition problem.
Especially, even if the universe is closed and likely to collapse just after the beginning of the universe, the potential energy of the inflaton easily dominates before the collapse and hence inflation naturally takes place~\cite{Linde:2005ht}.

Chaotic inflation models have been extensively studied in the framework of supergravity (SUGRA).
Chaotic inflation in SUGRA was realized in Ref.~\cite{Kawasaki:2000yn} by an $R$ symmetry and a shift symmetry.
The inflaton potential is given by small soft breaking of the shift symmetry in the superpotential $W$, which leads to the quadratic chaotic inflation model.

In Refs.~\cite{Kallosh:2010ug,Li:2013nfa,Harigaya:2014qza}, shift symmetry breaking was also introduced in the Kahler potential $K$,
with which the inflaton potential deviates from the quadratic one.
It was shown that the prediction on the spectral index and the tensor-to-scalar ratio is significantly altered.
As a result the model becomes consistent with the constraint from the Planck experiment~\cite{Harigaya:2014qza}.
Especially,
the tensor-to-scalar ratio $r$ is much smaller than the prediction of the purely quadratic model
if the shift-symmetry breaking in the Kahler potential is non-negligible.

However, if the shift symmetry is largely broken in the Kahler potential, the inflaton potential obtains large SUGRA effects for large inflaton field values.
The inflaton potential depends on the inflaton field through an exponential factor $e^K$ and hence it blows up exponentially for large inflaton field values.%
\footnote{If the Kahler potential is logarithmic of the inflaton field, the exponential blow up is absent. We do not consider this case in this letter.}

It should be reminded that for chaotic inflation to naturally take place event in a closed universe, the inflaton potential must flat enough for a large field value where the inflaton potential energy is as large as the Planck scale~\cite{Linde:2005ht}.
If the inflaton potential is too steep for the large field value, inflation is unlikely to take place.
Thus, there is an upper bound on the magnitude of the shift symmetry breaking in the Kahler potential.

In this letter, we derive an upper bound on the magnitude of the shift symmetry breaking in the Kahler potential, using the framework proposed in Ref.~\cite{Harigaya:2014qza}.
The upper bound leads to the lower bound on the tensor-to-scalar ratio, $r\gsim 0.1$.

\section{Supergravity chaotic inflation}

In this section, we review a chaotic inflation model in SUGRA proposed in Ref.~\cite{Kawasaki:2000yn}.
In SUGRA, the scalar potential is provided by the Kahler potential $K(\phi^i,\phi^{\bar{i}^\dag})$ and the superpotential $W(\phi^i)$,
where $\phi^i$ and $\phi^{\bar{i}\dag}$ are chiral multiplets and their conjugate, respectively.%
\footnote{$D$ term potentials are irrelevant for our discussion.}
The scalar potential is given by
\begin{eqnarray}
\label{eq:potential}
V &=&  e^K \left[
K^{\bar{i}i}D_i W D_{\bar{i}}W^\dag - 3 |W|^2
\right],\nonumber\\
D_i W &\equiv& W_i + K_i W,
\end{eqnarray}
where subscripts $i$ and $\bar{i}$ denote derivatives with respect to $\phi^i$ and $\phi^{\bar{i}\dag}$.
$K^{\bar{i}i}$ is the inverse of the matrix $K_{i\bar{i}}$.
Here and hereafter, we take a unit with the reduced Planck scale $\Mpl \simeq 2.4\times 10^{18}$ GeV being unity.

Let us introduce two chiral multiplets $\Phi$ and $X$, and consider the following super and the Kahler potentials,
\begin{eqnarray}
W &=& m X \Phi,\nonumber\\
K &=& K\left(XX^\dag, \left(\Phi + \Phi^\dag\right)^2\right) = \frac{1}{2} (\Phi + \Phi^\dag)^2 + XX^\dag + \cdots,
\label{eq:superKahler}
\end{eqnarray}
where $\cdots$ denotes higher dimensional terms and $m$ is a dimensionful parameter.
The super and the Kahler potentials are generic under an $R$ symmetry, a $Z_2$ symmetry and a shift-symmetry, which are summarized in Table~\ref{tab:charge}.%
\footnote{Shift symmetry breaking terms such as $K \supset F \left(\left(m\Phi + m^* \Phi^\dag\right)^2\right)$ does not change the following discussion since $m$ is very small, $m=O(10^{-5})$.}
$m$ is a holomorphic suprion which expresses shift symmetry breaking.
We discuss a model without $Z_2$ symmetry later.

\begin{table}
\begin{center}
 \begin{tabular}{|c|c|c|c|c|}
\hline
 & $R$ & $Z_2$ & shift \\
\hline
 $X$ & $2$ & $-1$ & $X\rightarrow X$ \\
 $\Phi$ & $0$ & $-1$ & $\Phi \rightarrow \Phi + i c$\\
$m$ & $0$& $+1$& $m\rightarrow m \frac{\Phi}{\Phi + ic}$\\
\hline
 \end{tabular}
\caption{Charge assignment of (spurious) fields. $c$ is an arbitrary real number.}
\label{tab:charge}
\end{center}
\end{table}

Due to the shift symmetry, the imaginary part of the scalar component of $\Phi$, which we denote as $\phi /\sqrt{2}$, does not obtain the exponential factor in Eq.~(\ref{eq:potential}).
The potential of $\phi$ remains flat for $|\phi|\gg1$, and hence
we identify $\phi$ as the inflaton.
The potential of $\phi$ is given by
\begin{eqnarray}
V (\phi) = \frac{1}{2} m^2\phi^2,
\end{eqnarray}
where we have put $X = {\rm Re} (\Phi) =0$, assuming that they obtain positive Hubble induced masses by Planck scale suppressed interactions.
 
The holomorphic shift symmetry breaking parameter $m$ is determined by the observed magnitude of the curvature perturbation ${\cal P}_\zeta \simeq 2.2\times 10^{-9}$~\cite{Ade:2013zuv},
\begin{eqnarray}
m \simeq 6\times 10^{-6} = 2\times 10^{13}~{\rm GeV}.
\end{eqnarray}
The spectral index $n_s$ and the tensor-to-scalar ratio $r$ are given by
\begin{eqnarray}
n_s = 1-\frac{2}{N_e} \simeq 0.967~~(N_e = 60),\nonumber\\
r = \frac{8}{N_e} \simeq 0.133~~(N_e = 60),
\label{eq:prediction}
\end{eqnarray}
where $N_e$ is the number of e-foldings corresponding to the pivot scale.

In addition to the holomorphic shift symmetry breaking parameter $m$, we introduce a real shift symmetry breaking parameter ${\cal E}$~\cite{Harigaya:2014qza} whose transformation law under the shift symmetry is
\begin{eqnarray}
{\cal E} \rightarrow {\cal E}\frac{(\Phi-\Phi^\dag)^2}{(\Phi-\Phi^\dag + 2ic)^2}.
\end{eqnarray}
Kahler potential terms which break the shift symmetry are given by%
\footnote{
Shift symmetry breaking terms such as $K \supset  XX^\dag {\cal E} (\Phi-\Phi^\dag)^2$ also contributes to the scalar potential.
As long as the inflaton dynamics for $|{\cal E}\phi^2|\lsim 1$ is concerned,
the contribution can be absorbed by redefinitions of $c_{2n}$.
}
\begin{eqnarray}
\label{eq:Kahlerbreaking}
K &\supset & F \left(
{\cal E}\left( \Phi - \Phi^\dag\right)^2\right)\nonumber \\
&\equiv&  \sum_{n=1}^{\infty} \frac{c_{2n}}{(2n)!} {\cal E}^n (\Phi - \Phi^\dag)^{2n} \nonumber \\
&=&  \frac{{\cal E}}{2} c_2 (\Phi - \Phi^\dag)^2 + \frac{{\cal E}^2}{4!}c_4 (\Phi - \Phi^\dag)^4 + \cdots.
\end{eqnarray}
The scalar potential of the inflaton $\phi$ is now given by
\begin{eqnarray}
V(\phi) = \frac{1}{2} m^2 \phi^2 \times {\rm exp}\left[ \sum_{n=1}^{\infty} \frac{ 2^n c_{2n}}{(2n)!} (-{\cal  E})^n \phi^{2n} \right].
\end{eqnarray}
As we will see, the inflaton dynamics for large field values with $|{\cal E}\phi^2|>1$ is relevant for the initial condition problem.
Hence, we assume that $|c_{2n}|= O(1)$, which ensures the convergence of the expansion in Eq.~(\ref{eq:Kahlerbreaking}) for $|{\cal E}\phi^2|>1$.%
\footnote{This is our basic assumption. If this is not satisfied, we cannot obtain any numerical constraints on model parameters
for chaotic inflation to naturally take place.}

Let us discuss the dynamics of the inflaton.
For $|{\cal E} \phi^2| \ll 1$, the first and second slow-roll parameters are given by~\cite{Harigaya:2014qza}
\begin{eqnarray}
\epsilon(\phi) =& \frac{1}{2}\left(\frac{V_\phi}{V}\right)^2 & = \frac{2}{\phi^2}
\left(
1 - 2 c_2{\cal E} \phi^2 + \frac{3c_2^2+2 c_4}{3}{\cal E}^2 \phi^4
\right) + O({\cal E}^3),\nonumber\\
\eta(\phi) =& \frac{V_{\phi\phi}}{V} & = \frac{2}{\phi^2}
\left(
1 - 5 c_2{\cal E} \phi^2 +\frac{6c_2^2+7c_4}{3} {\cal E}^2 \phi^4 \right) + O({\cal E}^3),
\end{eqnarray}
where subscripts $_\phi$ denote derivatives with respect to $\phi$.
The relation between the  number of e-foldings $N_e$ and the inflaton field value is
\begin{eqnarray}
N_e(\phi)
=
\frac{1}{4}(\phi^2-\phi^2_{\rm end}) +\frac{c_2}{8}{\cal E}(\phi^4-\phi_{\rm end}^4) +  \frac{3c_2^2-c_4}{36}{\cal E}^2 (\phi^6 -\phi^6_{\rm end}) + O({\cal E}^3),
\end{eqnarray}
with $\phi_{\rm end}\simeq \sqrt{2}$.

\begin{figure}[tb]
\begin{center}
  \includegraphics[width=.6\linewidth]{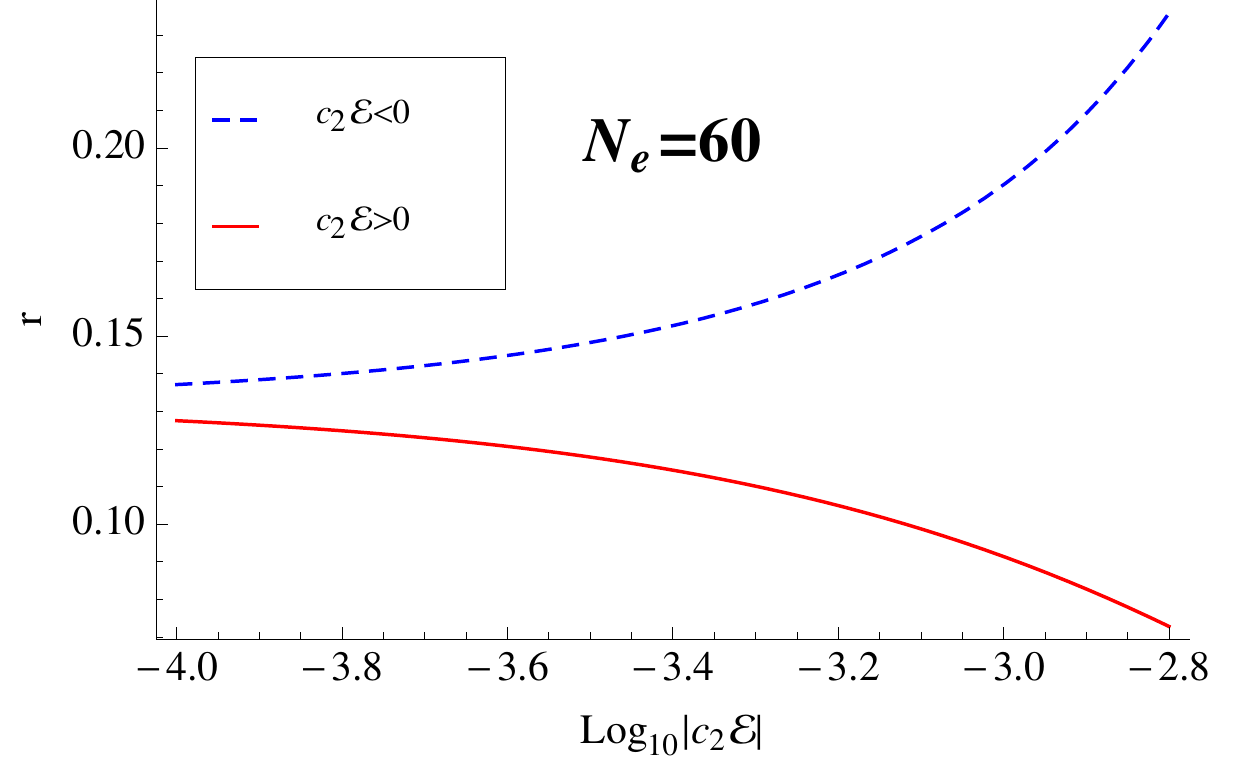}
 \end{center}
  \vspace{-1cm}
\caption{\sl \small
The tensor-to-scalar ratio $r$ as a function of the shift-symmetry breaking parameter ${\cal E}$.
}
\label{fig:fraction}
\end{figure}

The spectral index $n_s$ and the tensor-to-scalar ratio are written by the slow roll parameters as
\begin{eqnarray}
n_s = 1-6\epsilon + 2\eta,~~
r = 16\epsilon.
\end{eqnarray}
In Fig.~\ref{fig:fraction}, we show $r$ as a function of $c_2{\cal E}$.
It can be seen that $r$ can be smaller than the prediction of the purely quadratic model shown in Eq.~(\ref{eq:prediction}) for positive $c_2{\cal E}$. This is because the potential at around $N_e = 50\mathchar`-60 $ becomes flatter for positive $c_2{\cal E}$.
Since we are interested in the lower bound on $r$, we consider positive $c_2{\cal E}$ in the followings.

\section{Lower bound on tensor-to-scalar ratio}

In the previous section, we have reviewed the SUGRA chaotic inflation model.
We have shown that the tensor-to-scalar ratio can be smaller than the purely quadratic model by considering the real shift symmetry breaking spurion ${\cal E}$.
However, for ${\cal E} \neq 0$, the inflaton potential is steep for large field values due to the exponential factor in the scalar potential.

Note that for inflation to naturally take place even in a closed universe,
the inflaton potential must flat enough for a large inflaton field value where the inflaton potential energy is as large as the Planck scale~\cite{Linde:2005ht}.
Otherwise, the closed universe will collapses before slow-roll inflation starts.
Thus, there is an upper bound on the magnitude of the shift symmetry breaking ${\cal E}$,
if the universe is closed.

\begin{figure}[tb]
\begin{center}
  \includegraphics[width=.6\linewidth]{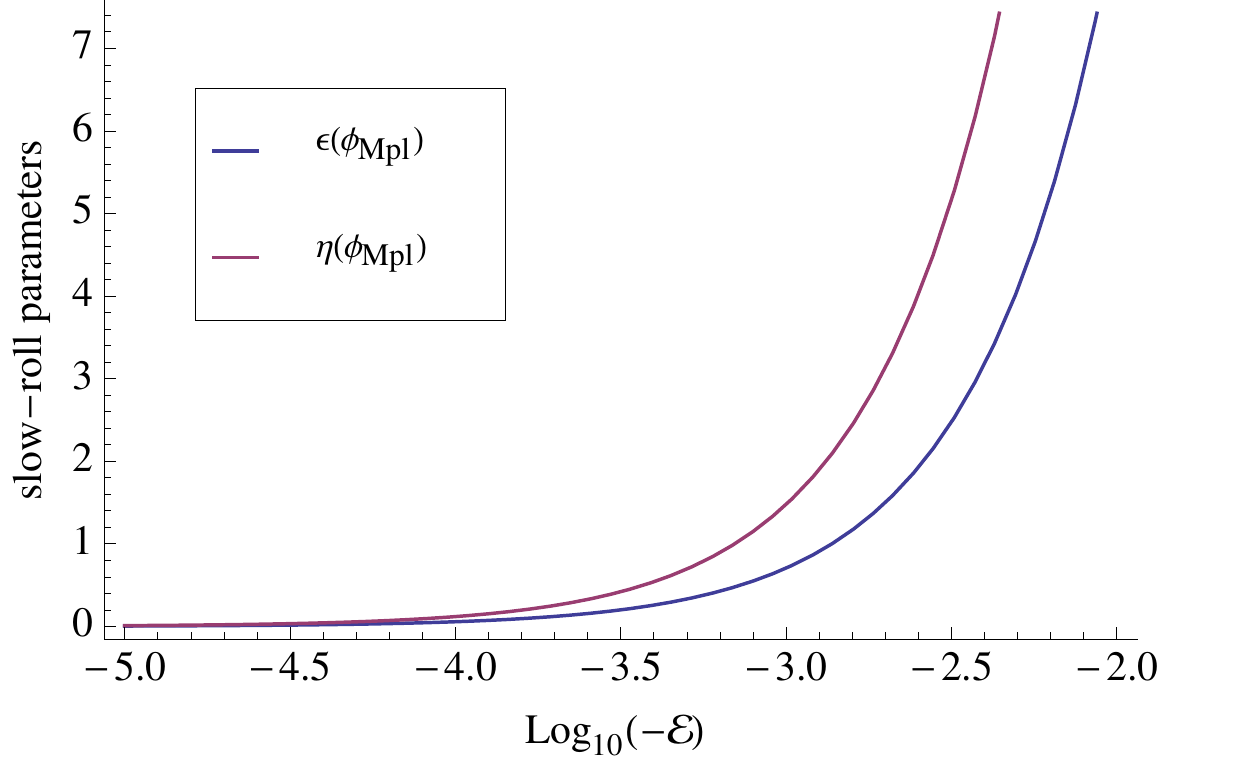}
   \vspace{-1cm}
 \end{center}
\caption{\sl \small
Slow roll parameters at the inflaton field value $\phi_{\Mpl}$ where the inflaton potential energy is as large as $\Mpl^4$.
}
\label{fig:slow roll}
\end{figure}

The upper bound can be roughly estimated as follows.
We define the inflaton field value $\phi_{\Mpl}$ as the field value where
\begin{eqnarray}
V (\phi_{\Mpl}) = 1~~ (= \Mpl^4).
\end{eqnarray}
Then the upper bound on ${\cal E}$ is estimated by requirements $\epsilon (\phi_{\Mpl}), |\eta (\phi_{\Mpl})|\lsim 1$.
In Fig.~\ref{fig:slow roll}, we show $\epsilon (\phi_{\Mpl})$ and $|\eta (\phi_{\Mpl})|$ as a function of ${\cal E}$.
We put $c_2 =-1$, $c_{2n} =1$ for $n>1$ and ${\cal E}<0$, for the time being.
We take negative ${\cal E}$ and positive $c_{2n} (n>1)$ in order to ensure a monotonously increasing inflaton potential,
and take negative $c_2$ to suppress the tensor-to-scalar ratio.
From Fig.~\ref{fig:slow roll}, we obtain the upper bound on $|{\cal E}|$, 
\begin{eqnarray}
\label{eq:upone}
|{\cal E}| \lsim 10^{-3},
\end{eqnarray}
which indicates the lower bound on $r$,
\begin{eqnarray}
\label{eq:lowonr}
r \gsim 0.1.
\end{eqnarray}

Let us confirm this estimation by solving the inflaton dynamics in a closed universe numerically.
For simplicity, we approximate the closed universe by a closed Friedmann-Robertson-Walker universe, whose metric is
\begin{eqnarray}
{\rm d}s^2 = - {\rm d}t^2 + a^2 (t)\left(\frac{{\rm d}r^2}{1-K r^2}  + r^2 {\rm d}\Omega^2\right),
\end{eqnarray}
with $K>0$. In the following we put $K=1$ by rescaling $r$.
After this rescaling the spatial curvature radius is given by $\sim a $, and hence a(0) is the initial physical size of the universe.
The Friedmann equation and the equation of motion of the inflaton are given by
\begin{eqnarray}
\label{eq:friedmann}
H^2 = \left(\frac{\dot{a}}{a}\right)^2 = \frac{\rho_{\rm tot}}{3} - \frac{1}{a^2},
\end{eqnarray}
\vspace{-1cm}
\begin{eqnarray}
\label{eq:eom}
\ddot{\phi} + 3  H \dot{\phi} + V_{\phi} =0,
\end{eqnarray}
where dots denote derivatives with respect to the time and $\rho_{\rm tot}$ is the total energy density of the universe.

We consider the evolution of the potential energy of the inflaton $\rho_{\rm pot} = V (\phi)$, the kinetic energy of the inflaton $\rho_{\rm kin} = \dot{\phi}^2/2$ and the spatial curvature density $\rho_K = -3 /a^2$.
The radiation and matter energy densities are negligible, because they are diluted much faster than the spatial curvature density is.
The gradient energy density is diluted as fast as the spatial curvature density is, and hence its effect on the inflaton dynamics can be represented by changing the initial spacial curvature density.

\begin{figure}[tb]
\begin{center}
  \includegraphics[width=.6\linewidth]{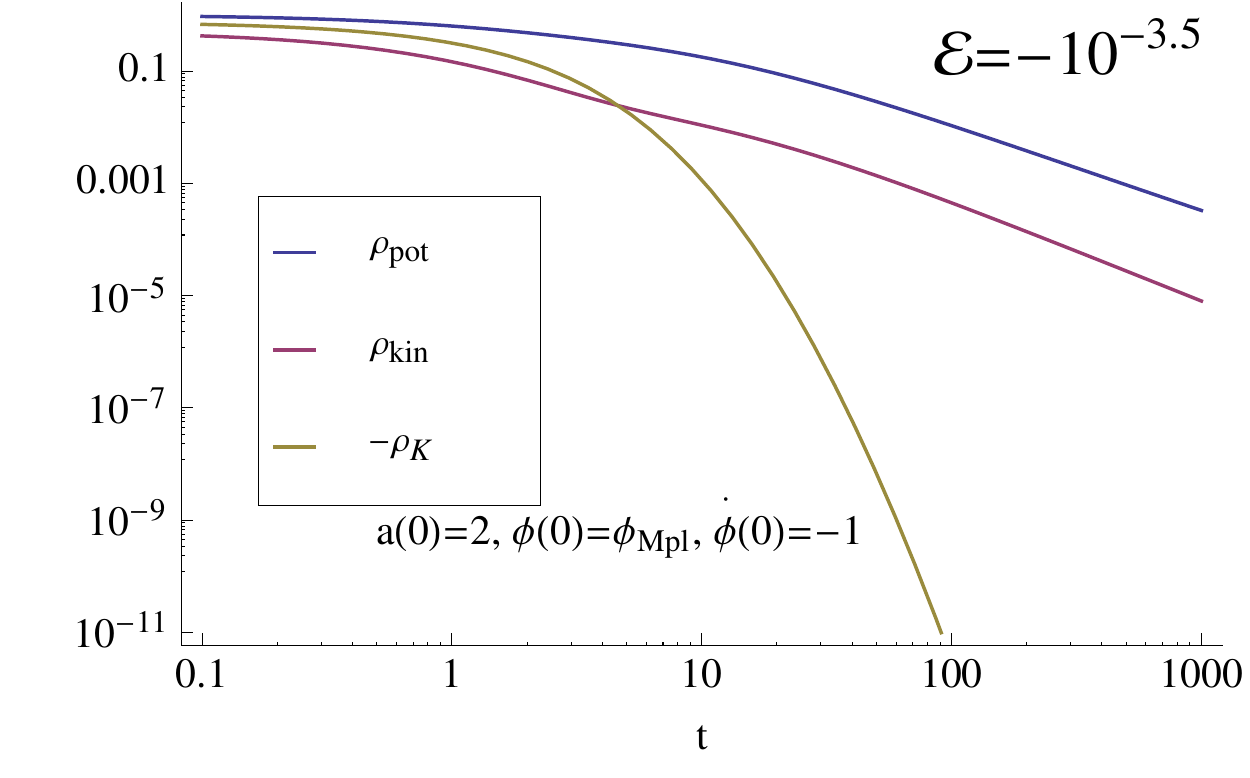}
 \end{center}
 \vspace{-1cm}
\caption{\sl \small
Evolution of  $\rho_{\rm pot}$, $\rho_{\rm kin}$ and $\rho_K $ for ${\cal E}=-10^{-3.5}$.
}
\label{fig:slow}
 \vspace{0.5cm}
\begin{center}
  \includegraphics[width=.6\linewidth]{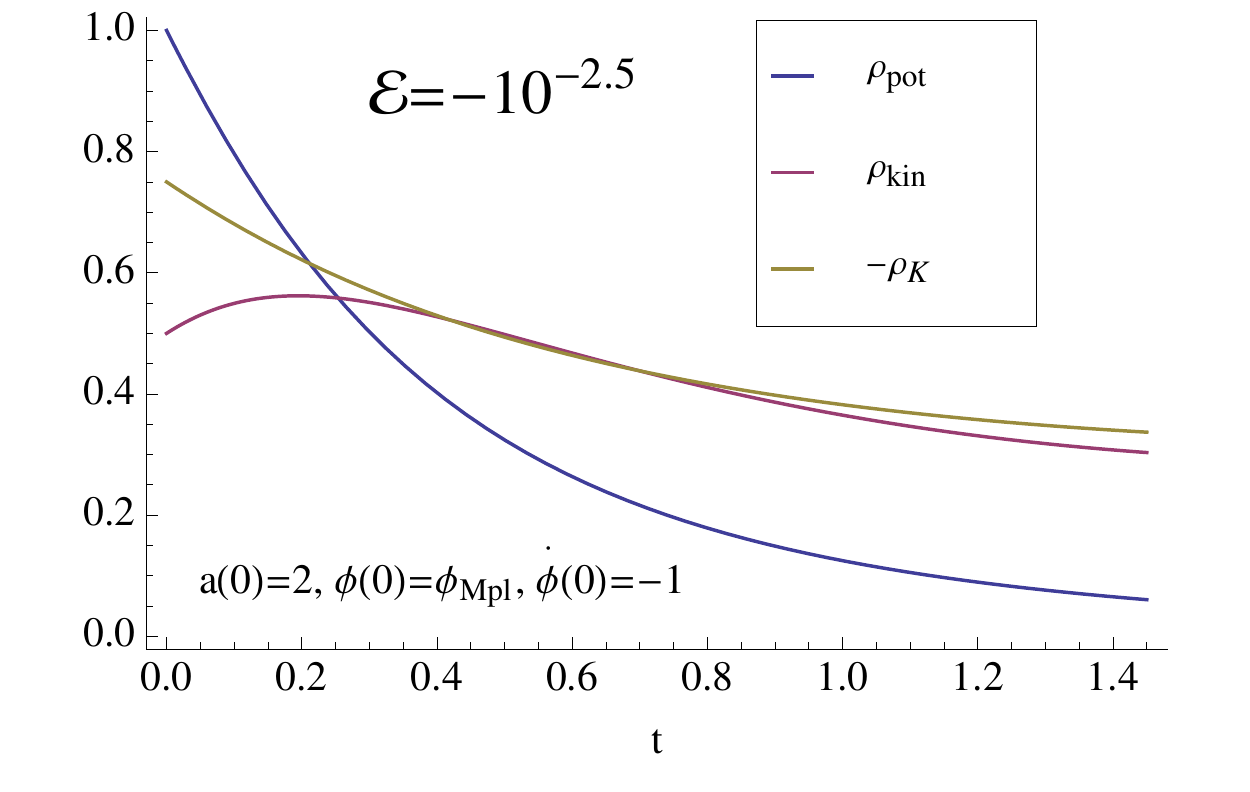}
 \end{center}
  \vspace{-1cm}
\caption{\sl \small
Evolution of  $\rho_{\rm pot}$, $\rho_{\rm kin} $ and $\rho_K $ for ${\cal E}=-10^{-2.5}$.
}
\label{fig:fast}
\end{figure}

In Fig.~\ref{fig:slow}, we show the evolution of $\rho_{\rm pot}$, $\rho_{\rm kin}$ and $\rho_K$ for ${\cal E} = - 10^{-3.5}$.
Here, we put the initial scale factor $a(0) =2$, $\phi(0)=\phi_{\Mpl}$ and $\dot{\phi}(0)=-1$.
Since slow-roll conditions are satisfied, the slow-roll inflation starts before the universe collapses.
In Fig.~\ref{fig:fast}, we show the same plot for ${\cal E} = -10^{-2.5}$.
The inflaton rolls down rapidly and hence the universe eventually collapses.
(Notice that the universe begins to collapse when $ H^2= (\rho_{\rm pot}+\rho_{\rm kin}+\rho_{K})/3 =0$.)

\begin{figure}[tb]
\begin{center}
  \includegraphics[width=.6\linewidth]{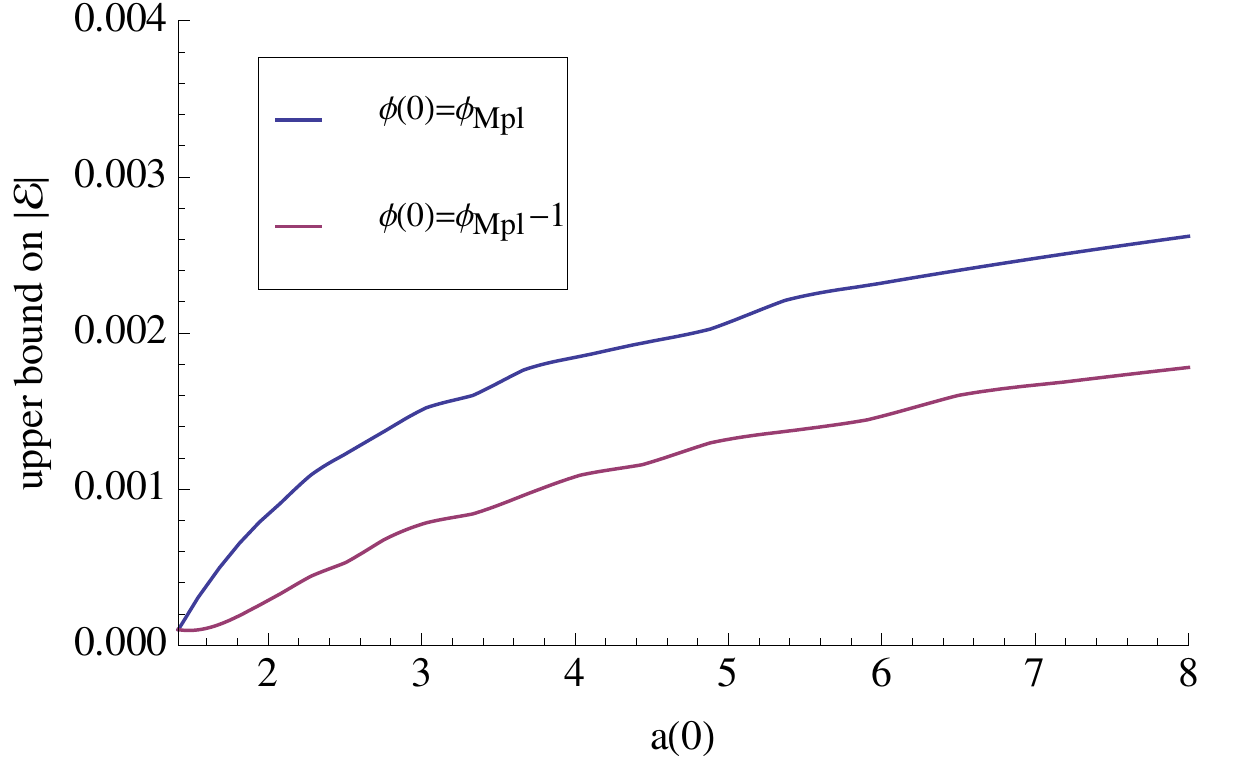}
 \end{center}
  \vspace{-1cm}
\caption{\sl \small
Upper bound on ${\cal E}$ for given $a(0)$ and $\phi(0)=\phi_{\Mpl},~\phi_{\Mpl}-1$.
}
\label{fig:upe}
\end{figure}

In Fig.~\ref{fig:upe}, we show the upper bound on $|{\cal E}|$ so that inflation begins before the universe collapses for various $a(0)$.
Here, we consider two cases with initial conditions $\phi(0)=\phi_{\Mpl}$ and $\phi(0)=\phi_{\Mpl}-1$ in order to estimate the effect of the fluctuation of the initial inflaton field value, which we expect to be as large as the Planck scale.
For larger $a(0)$, the initial size of the closed universe is larger and hence inflation is more likely to take place before the universe collapses.
Unless the initial size of the universe is much larger than the Planck length, $|{\cal E}|$ must be smaller than $10^{-3}$.

\begin{figure}[tb]
\begin{center}
  \includegraphics[width=.6\linewidth]{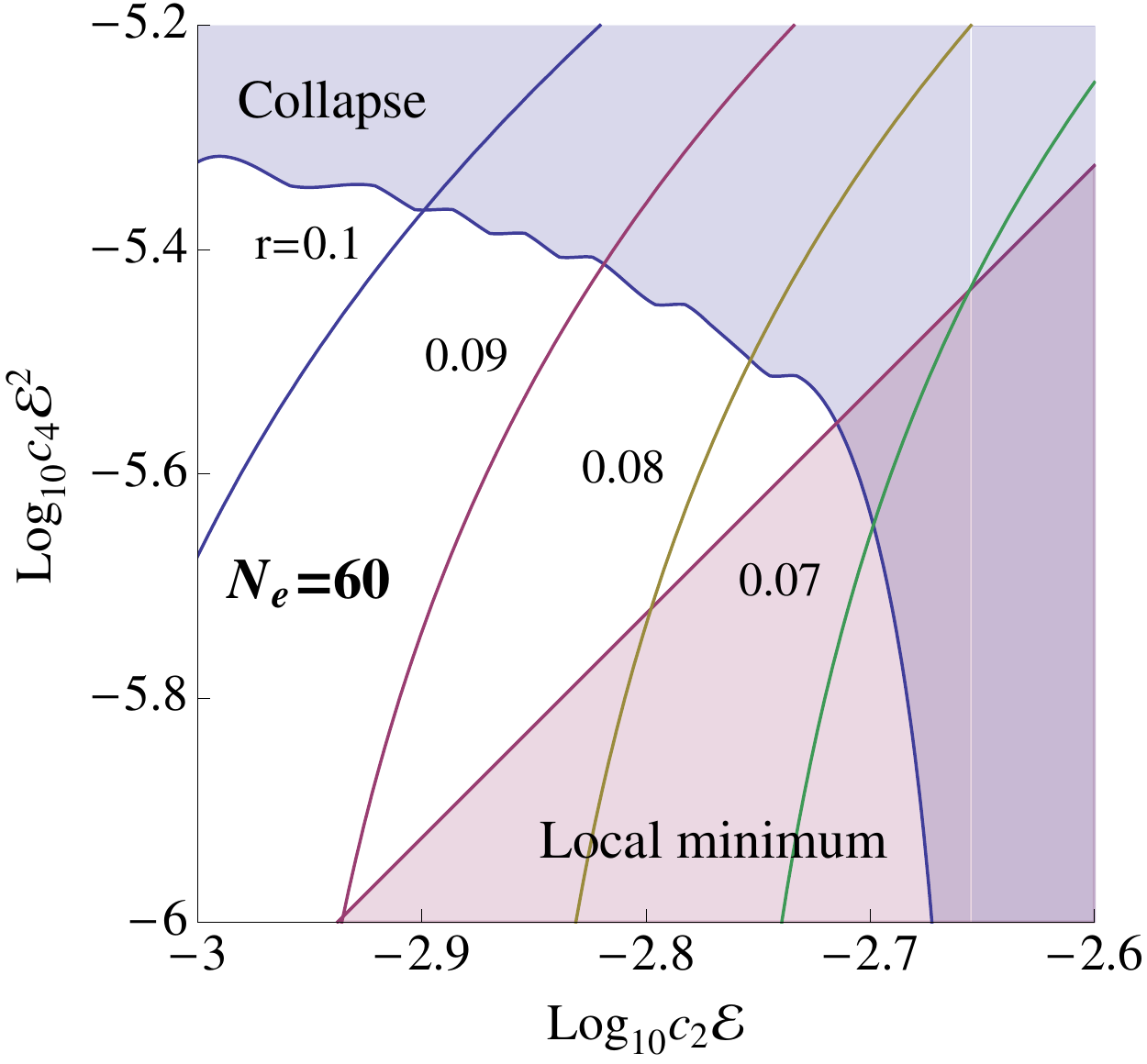}
 \end{center}
  \vspace{-1cm}
\caption{\sl \small
Bound on $c_2 {\cal E}$ and $c_4 {\cal E}^2$ for $a(0)=2$, $\phi(0)=\phi_{\Mpl}$ and $\dot{\phi}(0)=-1$.
}
\label{fig:c2c4}
\end{figure}

In the above analysis, we have assumed $c_2 = -1$ and $c_{2n}=1$ for $n>1$.
In fact, $O({\cal E}^3)$ terms in Eq.~(\ref{eq:Kahlerbreaking}) is unimportant for the inflaton dynamics.
(Notice that $V(\phi)\lsim 1$ requires $|{\cal E}\phi^2| \lsim 10$.)
In Fig.~\ref{fig:c2c4}, we show the bound on $c_2 {\cal E}$ and $c_4 {\cal E}^2$ for $a(0)=2$, $\phi(0)=\phi_{\Mpl}$ and $\dot{\phi}(0)=-1$ with neglecting $O({\cal E}^3)$ terms in Eq.~(\ref{eq:Kahlerbreaking}).
We also show the contour of $r$.
In the blue-shaded region, the closed universe collapses before inflation begins.
In the red-shaded region, there is a local minimum in the inflaton potential for $\phi<\phi_{\Mpl}$ and hence inflation ends only through quantum processes.%
\footnote{Around the upper edge of the red-shaded region, the field value of the inflaton after the quantum process is large enough that the e-findings of $50\mathchar`-60$ is possible afterwards.}
From Fig.~\ref{fig:c2c4}, we obtain the lower bound on $r$, $r\gsim 0.07$,
as expected in our rough estimation in Eq~(\ref{eq:lowonr}).

We have assumed the $Z_2$ symmetry in Eqs.~(\ref{eq:superKahler}) and (\ref{eq:Kahlerbreaking}) in the above model.
For a model without the $Z_2$ symmetry, we have, nevertheless, obtained a similar bound, $r\gsim 0.1$.

\section{Summary and discussion}
In this letter, we have considered the initial condition problem in a SUGRA chaotic inflation model with shift symmetry breaking in the Kahler potential.
We have estimated an upper bound on the shift symmetry breaking in the Kahler potential for a closed universe of a size of the Planck length not to collapse before inflation takes place.
The upper bound gives a lower bound on the tensor-to-scalar ratio, $r\gsim 0.1$.
If a smaller tensor-to-scalar ratio is observed, it indicates that our universe has its initial size larger than the Planck length, or it is open.

In the purely quadratic chaotic inflation model, the initial field value of the inflaton must be as large as $m^{-1}\sim 10^5$ in order for inflation to take place in a closed universe.
With shift symmetry breaking in the Kahler potential ${\cal E} = O(10^{-3})$, we see that the initial field value of the inflaton can be as small as $O(100)$.
Namely, the required range of the inflaton field value can be much smaller if the shift symmetry breaking in the Kahler potential is as large as ${\cal E}=O(10^{-3})$.
It is well known that in the purely quadratic chaotic inflation model, the stochastic effect dominates over the classical motion for large inflaton field values and hence eternal inflation necessarily occurs~\cite{Steinhardt:1983,Vilenkin:1983xq,Linde:1986fc}.
For the large shift symmetry breaking in the Kahler potential such as ${\cal E}=O(10^{-3})$, the stochastic effect is negligible because of the steep potential.
Eternal inflation does not take place.

\section*{Acknowledgments}
This work is supported by Grant-in-Aid for Scientific research from the
Ministry of Education, Science, Sports, and Culture (MEXT), Japan,
No. 25400248 (M.K.),
No.\ 26104009 and 26287039 (T.\,T.\,Y.),
and also by World Premier International Research Center Initiative (WPI Initiative), MEXT, Japan.
The work of K.H. is supported in part by a JSPS Research Fellowships for Young Scientists.


\begin{thebibliography}{99}


\bibitem{Guth:1980zm} 
  A.~H.~Guth,
  Phys.\ Rev.\ D {\bf 23}, 347 (1981).

\bibitem{Kazanas:1980tx} 
  D.~Kazanas,
  Astrophys.\ J.\  {\bf 241}, L59 (1980).


\bibitem{Linde:1981mu} 
  A.~D.~Linde,
  Phys.\ Lett.\ B {\bf 108}, 389 (1982).

\bibitem{Albrecht:1982wi} 
  A.~Albrecht and P.~J.~Steinhardt,
  Phys.\ Rev.\ Lett.\  {\bf 48}, 1220 (1982).

\bibitem{Starobinsky:1980te} 
  A.~A.~Starobinsky,
  Phys.\ Lett.\ B {\bf 91}, 99 (1980).


\bibitem{Mukhanov:1981xt} 
  V.~F.~Mukhanov and G.~V.~Chibisov,
  JETP Lett.\  {\bf 33}, 532 (1981)
  [Pisma Zh.\ Eksp.\ Teor.\ Fiz.\  {\bf 33}, 549 (1981)];
%
\bibitem{Hawking:1982cz} 
  S.~W.~Hawking,
  Phys.\ Lett.\ B {\bf 115}, 295 (1982).
\bibitem{Starobinsky:1982ee} 
  A.~A.~Starobinsky,
  Phys.\ Lett.\ B {\bf 117}, 175 (1982).
\bibitem{Guth:1982ec} 
  A.~H.~Guth and S.~Y.~Pi,
  Phys.\ Rev.\ Lett.\  {\bf 49}, 1110 (1982).
\bibitem{Bardeen:1983qw} 
  J.~M.~Bardeen, P.~J.~Steinhardt and M.~S.~Turner,
  Phys.\ Rev.\ D {\bf 28}, 679 (1983).


\bibitem{Hinshaw:2012aka} 
  G.~Hinshaw {\it et al.}  [WMAP Collaboration],
  Astrophys.\ J.\ Suppl.\  {\bf 208}, 19 (2013)
  [arXiv:1212.5226 [astro-ph.CO]].

\bibitem{Ade:2013zuv} 
  P.~A.~R.~Ade {\it et al.}  [Planck Collaboration],
  arXiv:1303.5076 [astro-ph.CO].

\bibitem{Linde:1983gd} 
  A.~D.~Linde,
  Phys.\ Lett.\ B {\bf 129}, 177 (1983).


\bibitem{Linde:2005ht} 
  A.~D.~Linde,
  Contemp.\ Concepts Phys.\  {\bf 5}, 1 (1990)
  [hep-th/0503203].



\bibitem{Kawasaki:2000yn} 
  M.~Kawasaki, M.~Yamaguchi and T.~Yanagida,
  Phys.\ Rev.\ Lett.\  {\bf 85}, 3572 (2000)
  [hep-ph/0004243].


\bibitem{Kallosh:2010ug}
  R.~Kallosh and A.~Linde,
  JCAP {\bf 1011}, 011 (2010)
  [arXiv:1008.3375 [hep-th]].


\bibitem{Li:2013nfa} 
  T.~Li, Z.~Li and D.~V.~Nanopoulos,
  JCAP {\bf 1402}, 028 (2014)
  [arXiv:1311.6770 [hep-ph]].

\bibitem{Harigaya:2014qza} 
  K.~Harigaya and T.~T.~Yanagida,
  Phys.\ Lett.\ B {\bf 734}, 13 (2014)
  [arXiv:1403.4729 [hep-ph]].

\bibitem{Steinhardt:1983}
Steinhardt, P J 1983 ``Natural inflation," in The Very Early Universe, Proceedings of the Nuffield Workshop, Cambridge, 21 June-9 July, 1982, eds: Gibbons, G W, Hawking, S W and Siklos, S T C (Cambridge: Cambridge University Press), pp. 251-66.

\bibitem{Vilenkin:1983xq} 
  A.~Vilenkin,
  Phys.\ Rev.\ D {\bf 27}, 2848 (1983).

\bibitem{Linde:1986fc} 
  A.~D.~Linde,
  Mod.\ Phys.\ Lett.\ A {\bf 1}, 81 (1986);
  Phys.\ Lett.\ B {\bf 175}, 395 (1986).

\end{thebibliography}
\end{document}